\documentclass[aps,12pt,superscriptaddress,amsfonts,amssymb,amsmath]{revtex4}
\pdfoutput=1

\usepackage{graphicx}
\usepackage{epsfig}
\usepackage{makeidx}
\usepackage{epstopdf}
\usepackage{natbib}
\usepackage{xcolor}
\usepackage{subcaption}
\usepackage{amsmath}
\usepackage{appendix}

\begin{document}

\title{Simple circuit and experimental proposal for the detection of gauge-waves}

\author{F.\ Minotti \footnote{Email address: minotti@df.uba.ar}}
\affiliation{Universidad de Buenos Aires, Facultad de Ciencias Exactas y Naturales, Departamento de F\'{\i}sica, Buenos Aires, Argentina}
\affiliation{CONICET-Universidad de Buenos Aires, Instituto de F\'{\i}sica Interdisciplinaria y Aplicada (INFINA), Buenos Aires, Argentina}

\author{G.\ Modanese \footnote{Email address: giovanni.modanese@unibz.it}}
\affiliation{Free University of Bozen-Bolzano \\ Faculty of Engineering \\ I-39100 Bolzano, Italy}

\linespread{0.9}

\begin{abstract}
Aharonov-Bohm electrodynamics predicts the existence of traveling waves of pure potentials, with zero electromagnetic fields, denoted as gauge waves, or g-waves for short. In general, these waves cannot be shielded by matter since their lack of electromagnetic fields prevents the material from reacting to them. However, a not-locally-conserved electric current present in the material does interact with the potentials in the wave, giving the possibility of its detection. In \cite{EPJC2023} the basic theoretical description of a detecting circuit was presented, based on a phenomenological theory of materials that can sustain not-locally-conserved electric currents. In the present work we discuss how that circuit can be built in practice, and used for the effective detection of g-waves.  
\end{abstract}

\maketitle

\section{Introduction}
\label{sec:introd}

The extended theory of electrodynamics by Aharonov and Bohm \cite{ohmura1956new,aharonov1963further,van2001generalisation,hively2012toward,Modanese2017MPLB,modanese2017electromagnetic,arbab2017extended,hively2019classical,Hively_2021,minotti2021quantum,minotti2022electromagnetic} is a generalization of the familiar Maxwell theory which allows to couple the e.m.\ field also to sources where charge is not locally conserved -- a phenomenon that is expected to occur in certain physical system with meso- or macroscopic quantum effects. Just because of this more general setting, the theory is not gauge invariant, and the electric and magnetic potentials $\phi$, $\mathbf{A}$ become univocally defined. 

The formal structure is simple. The potentials are computed through the equations ($\Box$ is the D'Alembert operator $c^{-2}\partial_t^2-\nabla^2$)
\begin{equation}
    \Box \phi=\frac{\rho}{\varepsilon_0}, \qquad \Box \mathbf{A}=\mu_0 \mathbf{j}
\label{eqs_potentials}
\end{equation}
(relativistic version: $\Box A^\mu \equiv \partial_\alpha \partial^\alpha A^\mu=\mu_0 j^\mu$, with $A^\mu=(c^{-1}\phi,\mathbf{A})$, $j^\mu=(c\rho,\mathbf{j})$).

With proper boundary conditions, the solutions of the eqs.\ (\ref{eqs_potentials}) are well-known retarded integrals, e.g.\ for $\phi$ we have
\begin{equation}
    \phi(\mathbf{x},t)=\frac{1}{\varepsilon_0} \int d^3y \frac{\rho(\mathbf{y},t-c^{-1}|\mathbf{x}-\mathbf{y}|)}{|\mathbf{x}-\mathbf{y}|}
\end{equation}
and similarly for $\mathbf{A}$.

The electric and magnetic fields $\mathbf{E}$, $\mathbf{B}$ are computed from the potentials as usual, namely $\mathbf{E}=\nabla \phi-\partial_t \mathbf{A}$, $\mathbf{B}=\nabla \times \mathbf{A}$ (relativistic form: $F_{\mu\nu}=\partial_\mu A_\nu-\partial_\nu A_\mu$).

The field equations for $\nabla \times \mathbf{E}$ and $\nabla \cdot \mathbf{B}$ are like the second and third Maxwell equations, while those for $\nabla \cdot \mathbf{E}$ and $\nabla \times \mathbf{B}$ contain additional terms depending on an ``auxiliary'' scalar field $S$ defined as $S=c^{-2}\partial_t \phi+\nabla \cdot \mathbf{A}$ (relativistic form: $S=\partial_\mu A^\mu$). More precisely, they are
\begin{eqnarray}
\nabla \cdot \mathbf{E} &=&\frac{\rho }{\varepsilon _{0}}-\frac{\partial S}{ 
\partial t}  \label{ABGauss} \\
\nabla \times \mathbf{B} &=&\frac{1}{c^2}\frac{\partial \mathbf{E} 
}{\partial t}+\mu _{0}\mathbf{j}+\nabla S
\label{ABAmpere}
\end{eqnarray}
(for their relativistic form see \cite{EPJC2023}).

The field $S$ satisfies the equation
\begin{equation}
    \Box S=\mu_0 I \equiv \mu_0 \left( \frac{\partial \rho}{\partial t}+\nabla \cdot \mathbf{j} \right)
\end{equation}
(relativistic version: $\Box S=\mu_0 I \equiv \mu_0 \partial_\mu j^\mu$), as follows immediately from its definition and from the equations for the potentials. We thus see that the source of $S$ is the quantity $I$, called ``extra-current'', which quantifies the violation of local conservation, being different from zero only in those regions where $\partial_t \rho \neq -\nabla \cdot \mathbf{j}$. If $I=0$ everywhere, then all equations reduce to their Maxwell form and $S$ can be set to zero through a gauge transformation.

Although the $S$ field has been called ``auxiliary'', because it is defined in terms of the potentials, it can manifest itself in principle in some small observable effects. Among these are waves with a longitudinal electric component and static currents with a partially  ``missing'' $\mathbf{B}$ \cite{minotti2021current,minotti2022electromagnetic}. The most relevant new physical effect predicted by the theory, however, is the possibility to detect pure gauge waves. Before explaining what these are, we need to complete the physical picture of the extended theory by giving the expression of the generalized Lorenz force exerted by fields and potentials on a probe. Let us call $\mathbf{f}$ the force density and $w$ the power density produced by the force (power supplied per unit volume). We have
\begin{eqnarray}
    \mathbf{f}&=&\rho_p \mathbf{E}+\mathbf{j}_p\times \mathbf{B}-I_p \mathbf{A} \label{force}\\
    w &=& \mathbf{j}_p\cdot \mathbf{E}-I_p \phi \label{power}
\end{eqnarray}
where $\rho_p$ and $\mathbf{j}_p$ are respectively the charge and current density in the probe, and $I_p$ the extra-current in the probe (by definition, $I_p$ is a density, i.e.\ referred to a unit volume). The two terms depending on $\mathbf{A}$ and $\phi$ are those which allow to measure directly the effects of the potentials. They vanish in the absence of extra-current, i.e.\ when the continuity relation $\partial_t \rho + \nabla \cdot \mathbf{j}=0$ (relativistic version: $\partial_\mu j^\mu=0$) holds everywhere in the probe.

\subsection{Gauge waves: properties and generation}

If we consider for simplicity the region where the wave can be considered as plane, a gauge wave takes the form
\begin{eqnarray}
    \phi(\mathbf{x},t) &=& \phi_0 e^{i(\omega t-\mathbf{k}\mathbf{x})} \\
    \mathbf{A}(\mathbf{x},t) &=& \mathbf{A}_0 e^{i(\omega t-\mathbf{k}\mathbf{x})}
\end{eqnarray}
in which it is understood that the real part of the right sides is taken, and where the amplitudes $\phi_0$, $\mathbf{A}_0$ satisfy the relation
\begin{equation}
    \omega \mathbf{A}_0=\mathbf{k} \phi_0
    \label{relaz-gw-1}
\end{equation}
This implies that the wave is longitudinal. It is a solution of the Aharonov-Bohm equations in vacuum, with standard relation between the wave vector $\mathbf{k}$ and frequency $\omega$, namely $\omega/|\mathbf{k}|=c$. It follows from (\ref{relaz-gw-1}) that $|\phi_0|=c|\mathbf{A}_0|$ and also that $\mathbf{k} \cdot \mathbf{A}_0=\frac{\omega}{c^2}\phi_0$. This means in turn that the potentials $\phi$ and $\mathbf{A}$ of a gauge wave satisfy the equation $\nabla \cdot \mathbf{A}+\frac{1}{c^2}\partial_t \phi=0$, also known in Maxwell electrodynamics as ``Lorentz gauge condition'' (relativistic form: $\partial_\mu A^\mu=0$).

Thus we see that in a gauge wave $S$ is zero by definition, and it is straightforward to check, using the relations which give $\mathbf{E}$ and $\mathbf{B}$ in terms of the potentials, that also $\mathbf{E}=0$ and $\mathbf{B}=0$ in the wave. This justifies the name of ``gauge waves''. It can also be verified using the generalized expressions for the field energy and momentum densities \cite{minotti2021quantum} that a gauge wave does not carry any energy or momentum. Actually, in a gauge-invariant theory such a wave would be regarded as non physical, because it is equivalent, via a gauge transformation, to one with potentials $\phi$ and $\mathbf{A}$ identically zero. In the Aharonov-Bohm theory, however, the potentials of a gauge wave are not equivalent to zero and cannot be ignored. They propagate without attenuation in normal media, but are able to interact with ``anomalous'' probes having $I\neq 0$, according to eqs.\ (\ref{force}), (\ref{power}).

The simplest method for generating gauge waves is through normal oscillating dipoles.
The physical A.-B.\ potentials of an oscillating dipole are (like in Maxwell theory in Lorentz gauge)
\begin{eqnarray}
    \phi(\mathbf{x},t) &=& \frac{\mu_0 c}{4\pi r} \dot{\mathbf{p}}(t-r/c)\cdot \mathbf{n} \\
    \mathbf{A}(\mathbf{x},t) &=& \frac{\mu_0}{4\pi r} \dot{\mathbf{p}}(t-r/c)
    \label{dip1}
\end{eqnarray}
where $r=|\mathbf{x}|$ and $\mathbf{n}=\mathbf{x}/r$ is the local propagation direction.

The vector potential $\mathbf{A}$ can be written as the sum of a longitudinal component
\begin{equation}
    \mathbf{A}_\mathbf{n}=(\mathbf{A}\cdot\mathbf{n})\mathbf{n}=\frac{\phi}{c} \mathbf{n}
\end{equation}
satisfying the gauge wave condition $|\mathbf{A}_\mathbf{n}|=\phi/c$, and a transverse component $\mathbf{A}_T$. 

Along the oscillation axis of the dipole we have that $\mathbf{A}_T=0$, so that the wave is longitudinal and a pure gauge wave. At any other point, the four-potential can be considered as the sum of a gauge wave $(\phi,\mathbf{A}_\mathbf{n})$ and a transverse vector potential $(0,\mathbf{A}_T)$. If the gauge wave encounters a normal conducting medium it is not affected (supposing the process is linear). The transverse component, on the other hand, generally interacts with the medium and in several cases can be strongly dampened, leaving as a result a pure gauge wave which propagates in the medium and then possibly gets out of the medium again into free space.

\subsection{Gauge waves: detection}

The general principle for computing the effect of a gauge wave on an anomalous conductor is the following \cite{EPJC2023}. Suppose that in the absence of the wave the conductor hosts a certain unperturbed current density $\mathbf{j}_u$ and unperturbed electric field $\mathbf{E}_u$. The electric power dissipated in the conductor is the volume integral of $\mathbf{j}_u \cdot \mathbf{E}_u$ and is supplied by an external generator which in the following we assume to be a DC generator.

In the presence of the gauge wave, the power $w$ supplied to the circuit does not change, because the wave does not carry any energy. The dissipated power, however, is given by the volume integral of $(\mathbf{j}\cdot \mathbf{E}-I\phi)$ (eq. (\ref{power}), omitting here the suffix ``$p$'' because it is clear that we are speaking of a probe). Therefore there must be a variation in the product $\mathbf{j}\cdot \mathbf{E}$:
\begin{equation}
    \delta (\mathbf{j}\cdot \mathbf{E})=\mathbf{j}\cdot \mathbf{E} -\mathbf{j}_u\cdot \mathbf{E}_u \simeq \delta \mathbf{j}\cdot \mathbf{E} + \mathbf{j}\cdot \delta \mathbf{E}
\end{equation}
and the integral of this quantity must be equal to the integral of $I\phi$. The products $\delta \mathbf{j}\cdot \mathbf{E}$ and $\mathbf{j}\cdot \delta \mathbf{E}$ can be expressed, for an ohmic conductor with resistance $R$, in terms of $R$ and of the variation of the current. About the integral of $I\phi$, it is found to depend on the gradient $\nabla\phi$ in the conductor, on the conductor length, on the current in it and on an adimensional constant $\gamma \ll 1$ which expresses in the simplest possible way the magnitude of the violation of local conservation associated to the extra-current $I$. In conclusion, oscillations of $\phi$ cause oscillations in the current.

For the details of the variation of the current in a specific device in response to $\nabla \phi$, see Sect.\ \ref{basic}.

The constant $\gamma$ depends on the material and is defined in the phenomenological ``$\gamma$-model'' \cite{EPJC2023}; a characteristic relation involving $\gamma$ is the generalized continuity equation $\partial_t \rho+(1+\gamma)\nabla \cdot \mathbf{j}=0$.

\subsection{Possible materials suitable for a gauge wave probe}

In Sect.\ \ref{basic} of this work we propose a detector of gauge waves based on the physical principles described above. The sensing element of the detector is a graphite ``antenna'' of the length of a few centimeters. The choice of graphite has two motivations.

The first motivation is theoretical and originates from some known properties of graphene, which can be seen as the elementary constituent of graphite. The electric conduction mechanism of graphene is peculiar and not yet fully understood. Conducting electrons behave as relativistic particles with null mass and a constant velocity equal to the Fermi velocity, described by chiral wavefunctions. Quantum tunneling in these materials becomes highly anisotropic and markedly different from the case of normal, non-relativistic electrons. It is in fact a condensed-matter realization of the Klein tunneling, which raises a paradox concerning local charge conservation \cite{katsnelson2006chiral,allain2011klein}. The solutions of the paradox that have been proposed go beyond single-particle wavefunctions and involve multi-particle quantum field theory processes (see e.g.\ \cite{alkhateeb2021relativistic} and refs.). However, it is well known that chiral symmetry is often broken at the quantum level.

On another front, there exist so-called ``first principles'' numerical calculations of conduction profiles in carbon macromolecules which reveal possible violations of local conservation. The first studies in this direction were due to J.\ Wang and collaborators \cite{li2008definition,zhang2011first}. They have shown that using a single-particle nonequilibrium Green's function technique coupled with the density-functional theory one obtains a conserved current density pattern only by adding to the density ``$\bar{\psi}\nabla\psi$'' derived from the molecular orbitals wavefunction a ``secondary'' current density extended in space which coincides with the term $\nabla S$ in the Aharonov-Bohm equations. This behavior is confirmed by calculations of Jensen, Garner, Solomon and collaborators for saturated chains of alkanes, silanes and germanes \cite{jensen2019current} and for linear carbon wires \cite{garner2019helical,garner2020three}. In \cite{jensen2019current} the authors apply the recipe by Wang et al.\ computing the secondary current via a Poisson equation. In \cite{garner2019helical} and \cite{garner2020three} they give detailed plots of the integrated local current density, as compared to the (constant) conserved current.

The second motivation is experimental and concerns directly graphite and its partial superconductive properties, observed even at room temperature \cite{esquinazi2014superconductivity,smith2015superconductivity,kopelevich2024global}. Although no generally accepted explanation of these superconductive properties exists, it is quite clear that conduction in graphite involves effects of macroscopic quantization. In this case one may expect some violations of local charge conservation because (1) several phenomenological non-BCS models of superconductivity comprise nonlocal wave equations \cite{modanese2018time,minotti2022electromagnetic}; (2) the number/phase uncertainty relation $\Delta N \Delta \phi \simeq 1$ and the non-commutativity of the quantum operators $\hat{\rho}$ and $\hat{\mathbf{j}}$ imposes a quantum uncertainty on the operator $\partial_t\hat{\rho}+\nabla\cdot\hat{\mathbf{j}}$ \cite{minotti2021quantum}.

\section{Basic circuit}
\label{basic}

\begin{figure*}[ht]
\includegraphics[width=10cm]{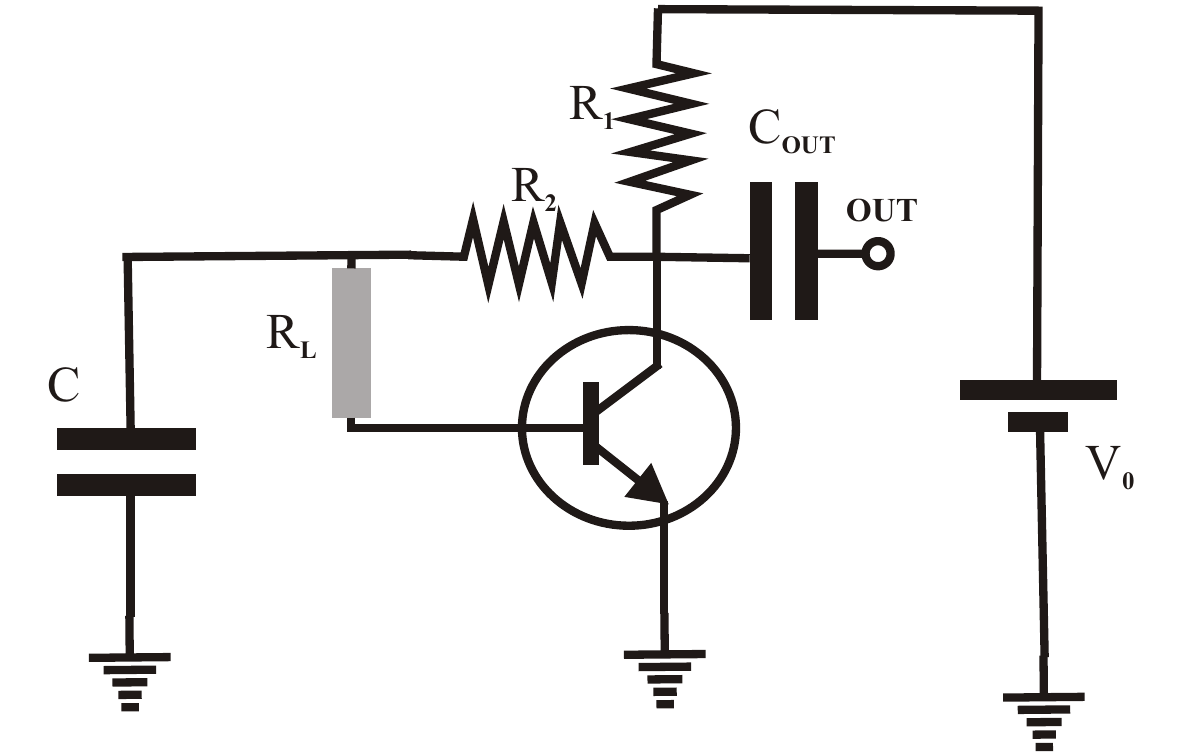}
\caption{Schematic of the detecting circuit.\label{QC1}}
\end{figure*}

The proposed circuit is shown in Fig \ref{QC1}

The circuit components are:

R$_{1}$ = 2.2 K$\Omega $,

R$_{2}$ = 10 K$\Omega $,

V$_{0}$ = 9 V,

C = 47 $\mu $F,

C$_{\text{OUT}}$ = 47 nF,

Transistor: 2N3904.

The element denoted as R$_{\text{L}}$ (that symbolizes its resistance $R_{L}$) is a graphite pencil lead of length of a few centimeters, that acts as the detecting element. The length is in fact related to the wavelength of the radiation to be detected. It should be sufficient for the value of $\Delta\phi$ of the gauge wave along the element to be non-negligible.  A practical limitation to the possible frequencies to be detected is given by the characteristics of the chosen transistor that indicate that it should work well up to a frequency of about 100 MHz.

\begin{figure*}[ht]
\includegraphics[width=10cm]{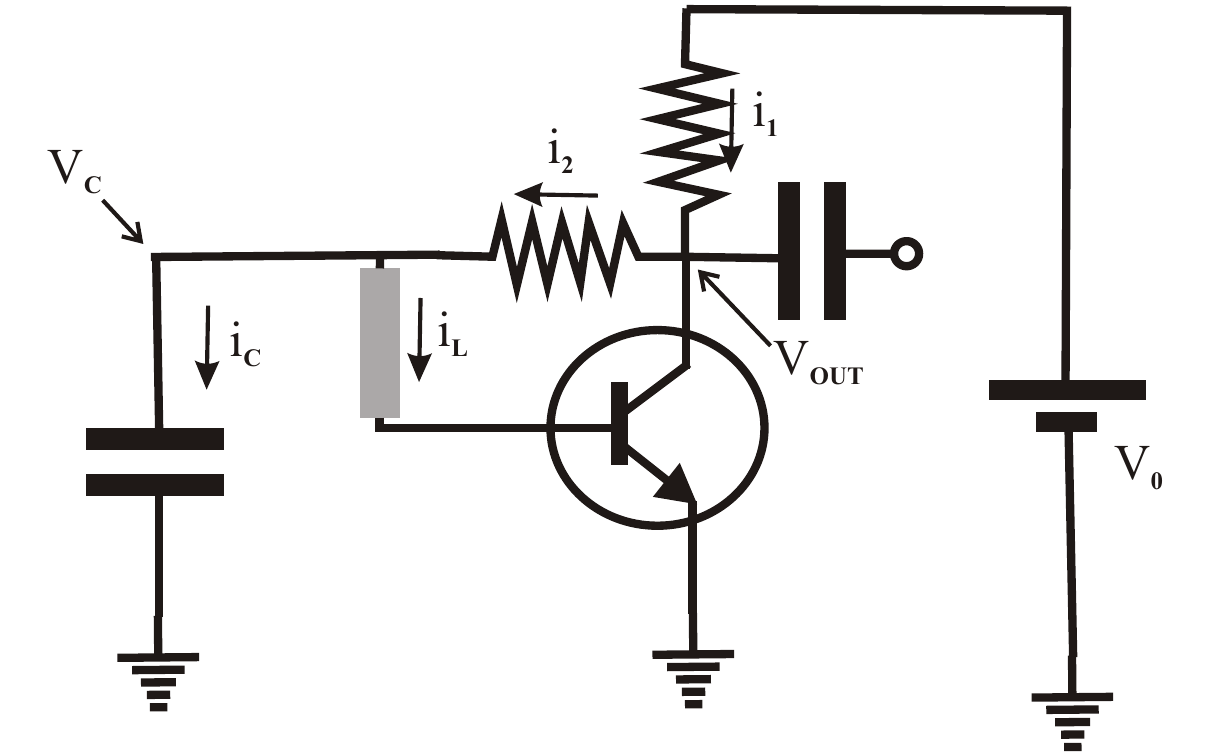}
\caption{Notation and conventions used in the circuit analysis.\label{QC2}}
\end{figure*}

In order to analyze the circuit behavior we refer to the Fig. \ref{QC2}

The following relations apply:
\begin{eqnarray*}
V_{OUT} &=&V_{0}-R_{1}i_{1}, \\
V_{C} &=&V_{0}-R_{1}i_{1}-R_{2}i_{2}, \\
i_{1}-i_{2} &=&\beta i_{L}.
\end{eqnarray*}
In the last relation $\beta $ is the amplification parameter of the
transistor, that relates the collector current ($i_{1}-i_{2}$) to the base
current ($i_{L}$). For the parameters chosen the transistor works in a
regime where $\beta \simeq 190$ (the value of $\beta $, however, cancels in
the final expressions).

For simplicity we assume the wave to be of sufficiently high frequency (to be quantified below) for the capacitor C to act as a short circuit, so that for the small-amplitude signal analysis we can
approximate $\delta V_{C}\simeq 0$. The full analysis should be done
for the Fourier amplitudes of the small signals, which would include the
impedance of C, and also $\delta i_{C}$ as an additional magnitude to be
determined (this is done in the next subsection). Note
besides that $\delta i_{C}=\left( 1+1/\beta \right) \delta i_{2}-\delta
i_{1}/\beta $ is limited by the resistors $R_{1}$ and $R_{2}$, so that it
cannot be arbitrarily large when the C impedance goes to zero, making the
assumption $\delta V_{C}\simeq 0$ possible.

In this way, from the second relation we have
\begin{equation}
\delta i_{2}=-\frac{R_{1}}{R_{2}}\delta i_{1}.  \label{di2}
\end{equation}

The time-varying signal detected through the capacitor C$_{\text{OUT}}$ is
given from the first relation as
\begin{equation*}
\delta V_{OUT}=-R_{1}\delta i_{1},
\end{equation*}
which using the third relation and (\ref{di2}) results in
\begin{equation}
\delta V_{OUT}=-\frac{R_{1}\beta }{1+R_{1}/R_{2}}\delta i_{L}.  \label{dVout}
\end{equation}

We can now determine the expected $\delta i_{L}$ considering that along R$_{\text{L}}$ there is a voltage $V_{L}=V_{C}-V_{BE}$, with $V_{BE}$ the voltage between base and emitter of about 0.7 V, whose variations are determined below. In this way we have that for the perturbation due to the gauge wave the energy conservation applied to the circuit gives:
\begin{equation}
\delta \left( i_{L}V_{L}\right) =\delta \left( R_{L}i_{L}^{2}\right) -\int
I\phi d^{3}x, \label{dilvl}
\end{equation}
where $I$ is the extra-current previously defined\cite{EPJC2023}.

Due to the assumed high frequencies we can again take $\delta V_{C}\simeq 0$, while $\delta V_{BE}$ has to be determined from the
transistor characteristics \cite{2N3904}. We thus have ($I_{B}$, and $I_{C}$ represent the base and collector currents)
\begin{equation*}
\delta V_{BE}=\left. \frac{\partial V_{BE}}{\partial I_{B}}\right)
_{I_{B}=i_{L}}\delta i_{L}=\beta \left. \frac{\partial V_{BE}}{\partial I_{C}
}\right) _{I_{B}=i_{L}}\delta i_{L}.
\end{equation*}
The working regime of the transistor for the proposed circuit corresponds to its DC operation at room temperature, with a collector DC current of about 3.5 mA, obtained solving the equations in the DC case, when $i_c=0$. For the small signal analysis around this regime we have
\begin{equation*}
\left. \frac{\partial V_{BE}}{\partial I_{C}}\right) _{I_{B}=i_{L}}\simeq
57\Omega ,
\end{equation*}
so that defining $R_{BE}\equiv \beta \left. \frac{\partial V_{BE}}{\partial
I_{C}}\right) _{I_{B}=i_{L}}$, we obtain
\begin{equation*}
\delta V_{L}=-R_{BE}\delta i_{L}.
\end{equation*}

From relation (\ref{dilvl}) we thus have, using also the $\gamma $-model developed in \cite{EPJC2023} to express $I$ in terms of the electric current, as detailed in that reference, 
\begin{eqnarray*}
V_{L}\delta i_{L}-R_{BE}i_{L}\delta i_{L} &=&2R_{L}i_{L}\delta i_{L}-\int
I\phi d^{3}x \\
&=&2R_{L}i_{L}\delta i_{L}-\gamma i_{L}\Delta \phi ,
\end{eqnarray*}
where $\Delta \phi $ is the difference of gauge-wave potential along R$_{
\text{L}}$. Since $R_{L}i_{L}=V_{L}$, we have
\begin{equation*}
\delta i_{L}=\gamma \frac{\Delta \phi }{R_{L}+R_{BE}},
\end{equation*}
giving finally from (\ref{dVout})
\begin{eqnarray}
\delta V_{OUT} &=&-\gamma \frac{R_{1}\beta }{\left( 1+R_{1}/R_{2}\right) }
\frac{\Delta \phi }{\left( R_{L}+R_{BE}\right) }  \notag \\
&=&-\gamma \frac{R_{1}}{\left( 1+R_{1}/R_{2}\right) }\frac{\Delta \phi }{
\left. \frac{\partial V_{BE}}{\partial I_{C}}\right) _{I_{B}=i_{L}}}.
\label{dVOUT_final}
\end{eqnarray}
where in the last line is was used that $R_{L}\ll R_{BE}$. For the values of
the parameters chosen we thus have
\begin{equation}
\delta V_{OUT}\simeq -31.6\gamma \Delta \phi .  \label{dVout_example}
\end{equation}
This voltage is definitely larger than one could obtain with the simpler passive circuit without transistor described in \cite{EPJC2023}, for which $\delta V_{OUT}\lessapprox \gamma \Delta \phi$.

\subsection{Inclusion of non-zero impedance for C}

If one does not assume that the impedance of C, $Z_{C}=\left(j \omega
C\right) ^{-1}$, is negligible ($j$ is the imaginary unit, and $\omega $ the
angular frequency of the Fourier component of the signal), one must add the relations
\begin{eqnarray*}
\delta i_{C} &=&\delta i_{2}-\delta i_{L}, \\
\delta V_{L} &=&\delta V_{C}-\delta V_{BE}=Z_{C}\delta i_{C}-R_{BE}\delta
i_{L}.
\end{eqnarray*}

Solving the complete system one readily obtains
\begin{equation*}
\delta V_{OUT}=-\frac{R_{1}\left[ \beta R_{2}+\left( 1+\beta \right) Z_{C}
\right] \gamma \Delta \phi }{R_{1}\left[ R_{L}+R_{BE}+\left( 1+\beta \right)
Z_{C}\right] +R_{2}\left( R_{L}+R_{BE}+Z_{C}\right) +Z_{C}\left(
R_{L}+R_{BE}\right) },
\end{equation*}
that reduces to the first line in (\ref{dVOUT_final}) when $Z_{C}\rightarrow
0$. For this approximation to be valid we see that the most stringent
condition is 
\begin{equation*}
\beta \left\vert Z_{C}\right\vert \ll R_{BE}=\beta \left. \frac{\partial
V_{BE}}{\partial I_{C}}\right) _{I_{B}=i_{L}},
\end{equation*}
that is
\begin{equation*}
\omega \gg \left[ C\left. \frac{\partial V_{BE}}{\partial I_{C}}\right)
_{I_{B}=i_{L}}\right] ^{-1},
\end{equation*}
which for the parameters considered corresponds to frequencies ($f=\left(
2\pi \right) ^{-1}\omega $) much higher than 60 Hz.

\section{Basic experimental proposal}

The circuit analyzed above is also sensitive to normal electromagnetic radiation, for
which the element R$_{\text{L}}$ acts as a normal antenna. It is thus
necessary to shield it from that radiation in order that only the gauge
component present interacts with R$_{\text{L}}$.

As considered in \cite{EPJC2023}, the far field, dipolar radiation of an emitting
antenna contains a gauge component with a scalar potential whose electric
field $-\nabla \phi $ has a magnitude similar to that of the ordinary
component (of course, the vector potential component of the gauge wave
generates a contribution $-\partial \mathbf{A}/\partial t$ that cancels that
of the scalar potential, yielding zero electromagnetic fields for the gauge wave). The
ratio of normal to gauge component depends on the orientation relative to
the emitting antenna (see \cite{EPJC2023} for details).

According to the $\gamma $-model, the detector is sensitive to the scalar
potential component if the cables connected to the graphite element R$_{\text{L}}$ are of different material (Copper, for instance), assumed to have a much lower value of $\gamma $.

In this way, if the circuit is conveniently shielded from normal electromagnetic radiation, only the gauge scalar component can be detected, because it cannot be shielded by virtue of its not interacting with matter. Any modulation of the incoming signal should thus be present in the circuit output.

To have an estimation of the expected magnitudes we use expression (\ref{dVout_example}), valid for the proposed circuit parameters. With $L\simeq 5.5$
cm, and for $\left\vert \nabla \phi \right\vert =1$ V m$^{-1}$ (corresponding to the gauge component of the electromagnetic field of a dipole antenna in the
region where the radiative flux is about $0.1\mu $W cm$^{-2}$) the expected amplitude of $\delta V_{OUT}$ is about $1.7\gamma $ V, so that even for $\gamma \sim 10^{-3}$ one has an easily detectable signal of the order of mV, whose modulation could be identifiable over the noise.

\section{Conclusions}
\label{conclusions}
In the present work we have presented a proposal for the practical implementation of a circuit for the detection of gauge-waves, a very special type of waves of pure potentials, whose detection is possible in the context of Aharonov-Bohm electrodynamics, but which are considered as non-physical (undetectable) in Maxwell electrodynamics. The rationale for the circuit operation is that a small fraction of the current that circulates in certain conductive materials does it in a non-conserved manner, giving rise to the so called ``extra-current'', which is shown in Aharonov-Bohm electrodynamics to directly interact with the scalar potential; in particular, with that component of the gauge wave. The circuit thus fulfills two purposes. On the one hand it establishes a controlled DC current in the detecting medium (a graphite element), whose supposedly non-conserved fraction interacts with the scalar potential component of the wave. On the other hand, the circuit amplifies the small current variations that result from that interaction, and which are detected as voltage variations easily analyzable on an oscilloscope. An apparently contradictory point concerning the conservation of energy is how a wave that does not transport energy is able to interact with a material medium. In fact, if the material is ``passive'', in the sense that in the absence of the gauge wave no macroscopic currents are present, the wave does not interact with the medium, according to the $\gamma$\ model used to describe media capable of supporting non-conserved currents \cite{EPJC2023}. On the other hand, in an ``active'' medium, the macroscopic current present is sustained by an external source, and it is this source the one that exchanges energy with the currents associated with the interaction with the wave. In a sense, the wave acts as a catalyst of the transduction of energy between the source and the current variations. Actually, the results here presented are derived from the conservation of energy relation valid in the context of Aharonov-Bohm electrodynamics \cite{minotti2021quantum}, so that the principle of conservation of energy is fulfilled. 
 
\bibliographystyle{ieeetr}
\bibliography{gwave}

\end{document}